\documentclass[twocolumn,showpacs,preprintnumbers,amsmath,amssymb,fleqn]{revtex4-2}

\usepackage{graphicx}
\usepackage{bm}
\usepackage{subfigure}
\usepackage{xcolor} 

\begin{document}

\title{Critical reexamination of a recent challenge to Bohmian mechanics}
\author{S. Di Matteo}
\email{sergio.dimatteo@univ-rennes.fr}
\affiliation{Univ Rennes, CNRS, IPR (Institut de Physique de Rennes) - UMR 6251, F-35000 Rennes, France}

\author{C. Mazzoli}
\affiliation{National Synchrotron Light Source II, Brookhaven National Laboratory, Upton, NY 11973, USA}

\date{\today}
\begin{abstract}
We reanalyze a recent experiment by Sharoglazova et al. [Nature (London) 643, 67 (2025)] highlighting the role of the transient regime. We prove that in the evanescent state of the stationary regime their experimental data can be interpreted in terms of Bohmian quantum mechanics. At the same time, Bohm’s quantum potential can be reinterpreted as a kinetic-energy term in the framework of Nelson’s stochastic quantum mechanics, with a hidden-variable, nonclassical speed fitting the experimental data as well. The experiment can be interpreted as well within orthodox quantum mechanics and is therefore not conclusive in selecting or challenging any framework.

\end{abstract}
\maketitle

\section{Introduction}

Recently, Sharoglazova {\it et al.} \cite{sharo} claimed to have proven a violation of the phase-velocity relation of Bohmian mechanics \cite{bohm}, $\vec{v}(\vec{r},t)=\frac{\vec{\nabla}S(\vec{r},t)}{m}$, by reporting the measurement of a quantity interpreted as a speed for an evanescent wave. The latter being real it implies $\vec{\nabla}S(\vec{r},t)=\vec{0}$, where $S(\vec{r},t)$ is the phase of the wave function $\psi(\vec{r},t)=\rho^{1/2}(\vec{r},t)e^{iS(\vec{r},t)/\hbar}$.

\begin{figure}[ht]
\centering
\includegraphics[width=0.45\textwidth]{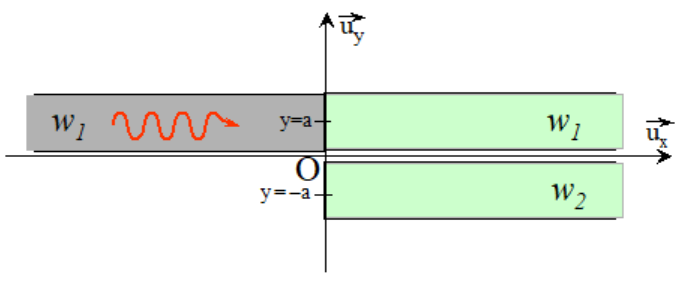}
\caption{Schematic view of the experimental setup of \cite{sharo}. The photon beam propagates along the waveguide $w_1$ from left to right. For $x\ge 0$, it finds a potential step $V_0$ and a second waveguide $w_2$, as described in the text. The coupling between waveguides leads to particle-density transfer from $w_1$ to $w_2$ (see Fig. \ref{fig2} for the three measured energy regimes).}
\label{fig1} 
\end{figure}

Actually, the basic idea behind their experimental claim had already been explained in their purely theoretical paper \cite{sharo2}, to which we mainly refer in our work: the authors realized a waveguide composed of several parts, schematically depicted in Fig. \ref{fig1}. In the challenging part of their experiment a propagating wave with energy $E<V_0$ enters the waveguide $w_1$ located around $y=+a$ from $x<0$. Then, it encounters the potential step $V_0$ existing for $x\ge 0$. The exponentially decaying wave experimentally measured in $w_1$ at $x>0$ is also found to fill in the second waveguide $w_2$ located around $y=-a$. The density profile in $w_2$ grows with $x$ for small $x$ (see also Fig. \ref{fig2}), so the reported measurements seem to imply that at $x\ge 0$ some density moves across the waveguides, from $w_1$ to $w_2$. The latter would result at odds with the phase-velocity relation of Bohmian mechanics for a real evanescent wave. This naturally leads to the challenge proposed by Sharoglazova {\it et al.}, who also quantified in terms of a speed the measured stationary density profiles. In their own words, the key hypothesis reads \cite{sharo}: 'If the probability amplitude is initially concentrated entirely in one of the states (here, $w_1$), the population in the initially unoccupied state (here, $w_2$) follows $\sin^2(J_0t)$ as a function of time', from which the deduction of the $w_1$ to $w_2$ motion at any time t.

However, the above argument is built around the assumption that, for $x \geq 0$, the presence of a density profile at time $t$ in any of the two waveguides is the necessary consequence of a current flow realized at that same time $t$. Yet, in order to displace density at $(x,y,t)$, we need $\partial_t\rho(x,y,t)\neq 0$, and therefore $\vec{\nabla}\cdot\vec{j}(x,y,t) \neq 0$ by the continuity equation, at odds with the stationary conditions $\vec{\nabla}\cdot\vec{j} = 0$ (or even with the condition $\vec{j}=\vec{0}$ in the evanescent state). 
This argument implies that either the measured density profile must be formed before that stationary condition is reached, i.e. in the transient regime, or that there is a permanent leakage in the experiment allowing $\vec{\nabla}\cdot\vec{j}(x,y,t) \neq 0$.
In this paper we develop the consequences of the former idea through the analysis of a nondissipative Hamiltonian dynamics equivalent to the theoretical model \cite{sharo2}. The latter one is discussed in \cite{drezet,drezet2}.

Moreover, it is interesting that in the description of \cite{sharo,sharo2} the key energy-parameter $|\Delta|$ is equal to the quantum potential, $|Q|$. The latter can be interpreted both as potential energy in terms of a static wave vector (not a speed) in Bohm's ontology, or as kinetic energy in terms of a nonclassical speed in Nelson's ontology \cite{nelson}. Thus, the experiment can be coherently interpreted in the framework of both ontologies, showing that none of them is either challenged or preferred.

To prove our main arguments as per above, the paper is organized in the following way:
Section II is devoted to the analysis of the transient regime in the geometry of Fig. \ref{fig1}, highlighting the role of the transversal current $j_y$ needed to obtain the density experimentally measured in the waveguide $w_2$ \cite{sharo}. 
In Sec. III we confirm that along the $x$ direction our full (2D) Hamiltonian provides the same results as the effective two-channel one-dimensional analysis performed in \cite{sharo2}.
Section IV is dedicated to the analysis of two nonorthodox ontologies of quantum mechanics (Bohm's and Nelson's interpretations), showing why the experiment can be explained in two different ways, and why each explanation is coherent (only) within the corresponding interpretation. A comparison with the 'orthodox' view is also offered, as well as a general discussion of our paper in the framework of the existing literature \cite{drezet,drezet2,waegell,matzkin,dickau}.  
Finally, in Sec. V we draw our conclusions highlighting the pedagogical potentials of this experimental setup. 
We have put in Appendix A the calculations leading to the stationary solution of the two-dimensional Hamiltonian (\ref{ham2D}), necessary to describe any transient transversal motion in the $y$ direction, whereas in Appendix B we recall the standard one-dimensional transient analysis \cite{review,moshinsky,moshinsky2}. We remark that the Hamiltonian (\ref{ham2D}) naturally brings us to the correct divergenceless conditions in the stationary regime, avoiding the introduction of exotic interpretations \cite{sharo3} and highlighting the importance of the transverse current $j_y$.


\section{Transient regime}

The experimental setup of \cite{sharo} consisted of two waveguides aligned along the $x$ direction, one centered at $y=+a$ ($w_1$), the other at $y=-a$ ($w_2$). The latter is limited to $x\ge 0$, as schematically shown in Fig. \ref{fig1}. For $x\ge 0$ a potential step $V_0$ is also added to both $w_1$ and $w_2$. The photons in this experimental setup 
behave like massive particles in two dimensions \cite{sharo}, and thus they can be modeled by the Schr\"odinger equation $i\hbar\partial_t\psi(x,y,t)= {\hat H}(x,y)\psi(x,y,t)$ with the following time-independent Hamiltonian in the $x$ and $y$ directions:

\begin{align}
\!\!\!\!\!\!{\hat H}(x,y) &= -\frac{\hbar^2}{2m}\left(\frac{\partial^2}{\partial_x^2} + \frac{\partial^2}{\partial_y^2}\right) + \theta(-x) \frac{1}{2}m\omega_0^2(y-a)^2 \nonumber \\
&+\theta(x) \left[\frac{1}{2}m\omega_0^2(|y|-a)^2 +V_0 \right]
\label{ham2D}
\end{align}

\noindent where $V_0$ is a positive constant representing the step potential at the origin, $\theta(x)$ is the Heaviside function (1 for $x>0$ and 0 for $x<0$), $m$ is the mass of the particle, and both waveguides have been modeled in the $y$ direction through harmonic potentials of angular frequency $\omega_0$ centered at $y=\pm a$. In what follows, we shall only consider the ground states of the harmonic potentials.
The stationary eigenstates of Hamiltonian (\ref{ham2D}) have the form $\psi(x,y)=\theta(-x)\psi_<(x,y)+\theta(x)\psi_>(x,y)$. The full stationary solution is given in the Appendix A for any energy. We now move to the transient solution in the evanescent energy region (see Fig. \ref{fig2} in Appendix A).

The reason to study the transient regime is simple. From the continuity equation, $\partial_t \rho(x,y,t) =-\vec{\nabla}\cdot\vec{j}$, integrating both terms in time from $t=0$ (when there is no density for $x>0$) to $+\infty$ (when the stationary state is reached), we get: 

\begin{align}
\rho(x,y)=-\int_0^{\infty}dt \left( \partial_x j_x +\partial_y j_y \right)
\label{divj}
\end{align}

\noindent where the time-independent $\rho(x,y)$ is the stationary density of particles. Of course, during this transient regime, the density has to move transversally from $w_1$ (where it enters at $x=0$, $t=0$) to the auxiliary waveguide $w_2$, so as to reproduce the stationary density at later times. This can only be done through a nonzero time-integral of $j_y$. 
We obtain (see Appendix B) the following expression for $j_y(x,y,t)$:

\begin{align} \label{jytrans}
\!\!\!\!\!\!\!\!\!\!\!\!\!\!\!\!\!\!\!\!\!\!\!\!\!\!\!\!\!\!\!\!\!\!\!\!\!\!\!\!\!\!\!\!\!\!\!\! &j_y(x,y,t)=\frac{\hbar}{m} Im[\psi^*(x,y,t)\partial_y\psi(x,y,t)] \\
&\!\!\!\!\!\!=\frac{-\hbar a}{4m\sigma^2}|T|^2e^{-\tfrac{q_2x}{2}}\chi_0^{(-)}\chi_0^{(+)} Im[{\rm erfc}^*(z_+){\rm erfc}(z_-)] \nonumber
\end{align}

\noindent where $\chi_0^{(\pm)}=\chi(y\pm a)$. We remark that for $t\rightarrow\infty$, as we reach the stationary state, the erfc becomes a real function and so $j_y$ is zero, as it should. Moreover, $j_y$ decreases longitudinally as $e^{-q_2x/2}$, and is initially negative, as $Im[{\rm erfc}^*(z_+){\rm erfc}(z_-)]$ is initially positive. Therefore the displacement is from $w_1$ to $w_2$. These are the main features of Eq. (\ref{jytrans}) in this specific case of sudden jump in the propagating pulse. Though a detailed analysis would require a precise knowledge of the initial wave packet, the main message of this section is robust: the density profile measured in both waveguides $w_1$ and $w_2$ in stationary conditions is formed by the nondivergenceless transient current $j_y(x,y,t)$ flowing {\it before} the stationary conditions are reached.

\section{Density measurement\\in stationary conditions}

We recall the operational measurement performed in \cite{sharo}, through a relative population analysis in stationary conditions. The relative density of the waveguide $w_2$ is compared to the full measured density as: 

\begin{align}
\rho_{a}(x)=\frac{|\psi_>(x,-a)|^2}{|\psi_>(x,a)|^2+|\psi_>(x,-a)|^2}
\label{rhoa}
\end{align}

\begin{figure}[t]
\centering
\includegraphics[width=0.50\textwidth]{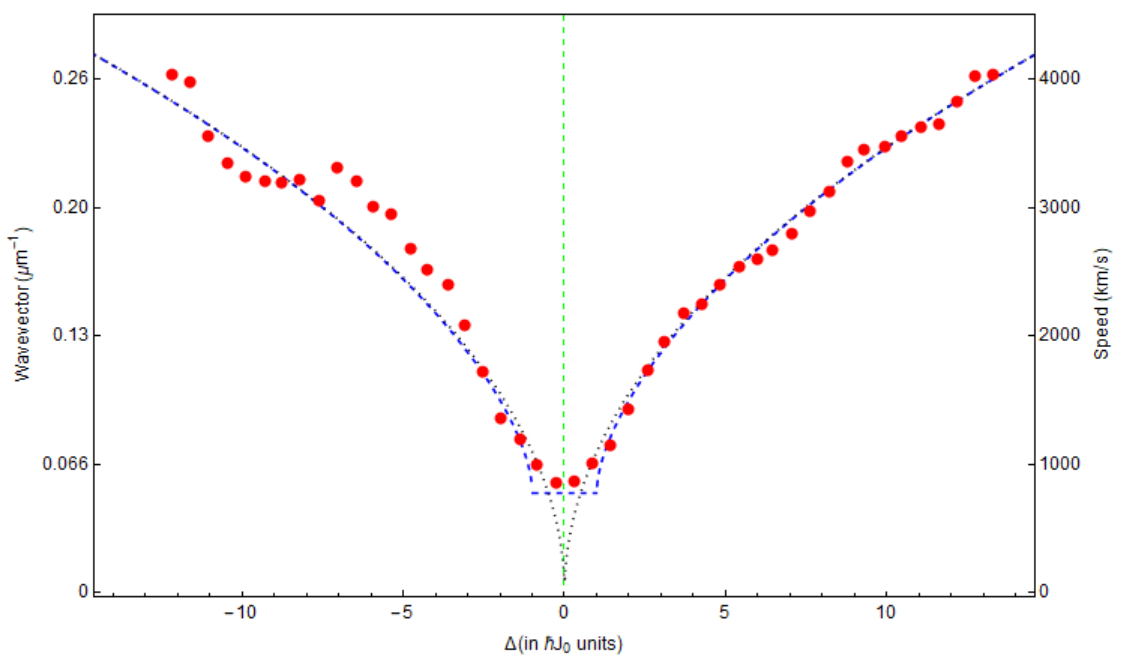} 
\caption{The stationary experimental data (red circles) from \cite{sharo} can be reproduced (dashed blue line) in the three energy regions by using Eq. (\ref{eqfig3}) as a function of $|\Delta|$ (in units $\hbar J_0$). Focusing on the evanescent energy region ($\tfrac{\Delta}{\hbar J_0}\le -1$), the dashed blue line can be interpreted in terms of a stationary wave vector $q_2$ (left scale of the ordinates). However, when $\tfrac{\Delta}{\hbar J_0}\ll -1$, it can also be interpreted as a nonclassical speed $u_x=\tfrac{\hbar q_2}{m}$ (right scale of the ordinates). See Sec. IV for the corresponding ontologies. The black dotted line correspond to the square-root model of \cite{sharo}.}
\label{fig3} 
\end{figure}

We can evaluate this quantity from Eq. (\ref{psievan}), neglecting $\chi_0(\pm 2a)\ll \chi_0(0)$. Through the above definitions, reminding that $k_1=\tfrac{mJ_0}{\hbar k_2}$ for any initial energy region (see classification in the Appendix A), we obtain the same dependencies as in \cite{sharo2}: 

\noindent a) Region 1 ($\tfrac{\Delta}{\hbar J_0}\ge 1$): 

$\rho_{a} = \sin^2(k_1x) \xrightarrow[k_1x\ll 1]{ } (k_1 x)^2= \left(\frac{mJ_0}{\hbar}\right)^2\frac{x^2}{k_2^2}$.

\noindent b) Region 2 ($\tfrac{|\Delta|}{\hbar J_0}\le 1$): 

$\rho_{a} = \frac{\cosh(q_-x)-cos(k_+x)}{2\cosh(q_-x)} \xrightarrow[q_x,k_+x\ll 1]{ } \frac{q_-^2+k_+^2}{4}x^2=\frac{mJ_0}{\hbar}x^2$. 

It can be rewritten as follows: $\rho_{a} = \left(\frac{mJ_0}{\hbar}\right)^2\frac{x^2}{k_J^2}$, with $k_J=\sqrt{\frac{mJ_0}{\hbar}}$, independent of $\Delta$.

\noindent c) Region 3 ($\tfrac{\Delta}{\hbar J_0}\le -1$): 

$\rho_{a} = \sinh^2(q_1x) \xrightarrow[q_1x\ll 1]{ }(q_1 x)^2= \left(\frac{mJ_0}{\hbar}\right)^2\frac{x^2}{q_2^2}$.

Rewriting $k_2$ and $q_2$ in terms of $\Delta$ in regions 1 and 3, respectively, we obtain: $\rho_a=\frac{mJ_0^2 x^2}{|\Delta|(1+\sqrt{1-(\hbar J_0/|\Delta|)^2})}$ (note the parabolic dependence on $x$ as fitted in \cite{sharo}). The dependence of $k_2$ (region 1), $k_J$ (region 2), and $q_2$ (region 3) in terms of $\Delta$ is therefore expressed as

\begin{equation}
\hspace{-1cm} q = \left\{ \begin{aligned}
&q_2 = \tfrac{1}{\hbar}\sqrt{m|\Delta|(1+\sqrt{1-(\hbar J_0/|\Delta|)^2})} && \text{if } \tfrac{\Delta}{\hbar J_0} \le -1 \\
&k_J = \sqrt{\frac{mJ_0}{\hbar}} && \text{if } \tfrac{|\Delta|}{\hbar J_0} \le +1 \\
&k_2 = \tfrac{1}{\hbar}\sqrt{m|\Delta|(1+\sqrt{1-(\hbar J_0/|\Delta|)^2})} && \text{if } \tfrac{\Delta}{\hbar J_0} \ge +1 
\end{aligned} \right.
\label{eqfig3}
\end{equation}

Equation (\ref{eqfig3}) is reproduced in Fig. \ref{fig3} (dashed blue line). It fits the experimental data well (see data availability section in [1]) without any need to invoke a speed: in fact, the primary information obtained from the measurements of $\rho_a$ in the stationary regime is that of a stationary wave vector. Such a wave vector was interpreted in \cite{sharo,sharo2} in terms of an associated speed. However, this is neither necessary nor trivial, as a speed with suitable characteristics (although profoundly different from the one proposed in \cite{sharo,sharo2}) would still be compatible with the interpretation of quantum mechanics (see Sec. IV B and IV D below). For completeness, we report in Fig. \ref{fig3} (black dotted line) also the energy-speed dispersion relation, $u_x=\sqrt{\tfrac{2|\Delta|}{m}}$, found in \cite{sharo,sharo2} by imposing the relation $u_x=\tfrac{\hbar q_2}{m}$.

In the next section, we shall clarify the relation between the two proposed interpretations of Fig. \ref{fig3}, the one in terms of static wave vector (Bohm's ontology) and the one in terms of speed (Nelson's ontology).


\section{Interpretation in Bohm's and Nelson's frameworks}

\subsection{Bohm's interpretation}

As it is well known \cite{bohm}, if we write the wave function $\psi(\vec{r},t)=R(\vec{r},t)e^{iS(\vec{r},t)/\hbar}$, the Schr\"odinger equation becomes a system of two coupled equations for the amplitude $R(\vec{r},t)$ and the phase $S(\vec{r},t)$: one is the continuity equation for a statistical ensemble of particles having density $\rho(\vec{r},t)=R^2(\vec{r},t)$ and speed $\vec{v}(\vec{r},t)=\frac{\vec{\nabla}S(\vec{r},t)}{m}$ and the other is the Hamilton-Jacobi equation corrected by a further term $Q(\vec{r},t)$:

\begin{align}
-\frac{\partial S}{\partial t} = \frac{(\vec{\nabla}S)^2}{2m} + V(\vec{r},t) + Q(\vec{r},t)
\label{HJ}
\end{align}

\noindent where the left-hand side term ($-\partial_t S$) represents the total energy, and the first two terms on the right-hand side are, respectively, the kinetic energy ($(\vec{\nabla}S)^2/(2m)$) and the potential energy, $V$.
Bohm \cite{bohm} interpreted the last term $Q(\vec{r},t)=-\tfrac{\hbar^2}{2m}\tfrac{\nabla^2R}{R}$ as potential energy of quantum-mechanical origin (it is the only term of the Hamilton-Jacobi equation where $\hbar$ is explicitly present). Depending on the shape of the real part of the wave function ($\tfrac{\nabla^2R}{R}$), such a term drives the motion of the particle together with the classical potential energy $E_p(\vec{r},t)$. In Bohm's ontology, it is at the origin of the peculiar behaviour of experimental findings in the quantum domain.

In the evanescent region 3, for $x>0$, the wave function is real, $S=0$ and therefore $Q(x>0,y)= E - V(x,y)= E_{k_0}-V_0 +\tfrac{1}{2}\hbar\omega_0-\tfrac{1}{2}m\omega_0^2(|y|-a)^2$. If we consider that all density measurements in the two waveguides have been performed for $y=\pm a$ and neglect the constant $\tfrac{1}{2}\hbar\omega_0$ of the zero-point energy harmonic oscillator (the experimental energy was measured with respect to $E_{k_0}$; see Appendix A), then $|Q(\pm a)|=|\Delta| = \tfrac{\hbar^2}{2m}(q_1^2+q_2^2)=\tfrac{\hbar^2}{2m}\left(\left(\tfrac{mJ_0}{\hbar q_2}\right)^2 +q_2^2\right)$, as plotted in Fig. \ref{fig3} (dashed blue line). 
The added value of this analysis resides in the identification of $|\Delta|$ with the quantum potential, so in Bohm's interpretation the measurement \cite{sharo} for $\tfrac{\Delta}{\hbar J_0} \le -1$ represents the $q_2$ dependence of the absolute value of the quantum potential. It is a static measurement and there is no need to introduce a speed: in Bohm's ontology the stability of the exponential spatial decay is due to the quantum force, determined by the quantum potential $Q$. Therefore, once the transient is finished, the density-transfer from one waveguide to the other stops and the stationary quantum potential, through Eq. (\ref{HJ}), forces the particle-ensemble to distribute according to the measured evanescent density. For $\Delta <0$, the higher $|\Delta|$, the higher the quantum force and, therefore, the higher $q_2$ measuring the curvature of the exponential decay.
This is the meaning of Fig. \ref{fig3} in the evanescent region 3, perfectly coherent within Bohm's ontology without introducing any speed.

\subsection{Nelson's interpretation}

However, we may have chosen to interpret Eq. (\ref{HJ}) with a different ontology: consider the following chain of identities on the 'quantum potential' term ($\rho=R^2$):

\begin{align}
Q&=-\frac{\hbar^2}{2m}\frac{\nabla^2R}{R}=-\frac{\hbar^2}{4m}\left[\frac{\nabla^2\rho}{\rho} -\frac{1}{2} \left(\frac{\vec{\nabla} \rho}{\rho} \right)^2 \right]\nonumber \\
& = \frac{1}{2} m \vec{u}\cdot\vec{u} - \frac{\hbar}{2}\vec{\nabla}\cdot\vec{u}
\end{align}

In this case we have described the same term $Q$ within the different ontology of Nelson's stochastic quantum mechanics \cite{nelson}, interpreting the energy term $Q$ \cite{note1} as a {\it nonclassical} speed, $\vec{u}=\tfrac{\hbar}{2m}\vec{\nabla}\ln\rho=+\tfrac{\hbar}{2m}\tfrac{\vec{\nabla}\rho}{\rho}$. The positive sign in the last expression has been explicitly written to underline the antidiffusive character of this velocity (it goes towards the density gradient instead of opposing it as in usual diffusion). Moreover, as in usual diffusion, $\vec{u}$ is time-reversal even (i.e., it does not change sign by reversing the time direction, as a propagative velocity would do).
In Nelson's interpretation, the total velocity of the quantum particle is given by the sum of two terms, $\vec{v}_{\rm tot}=\vec{v} +\vec{u}$. The first term is the usual center-of-mass velocity $\vec{v}=\tfrac{\vec{\nabla}S(\vec{r},t)}{m}$ and the second velocity $\vec{u}$, depending on the amplitude gradient, represents a fluctuating hidden-variable contribution (see also the different, less committed, interpretation given by \cite{hallreg}). It is worth noting that, from the definition $\vec{u}=\tfrac{\hbar}{2m}\tfrac{\vec{\nabla}\rho}{\rho}$, we get the nonzero current divergence: $\vec{\nabla}\cdot(\rho\vec{u})=\tfrac{\hbar}{2m}\nabla^2\rho$. However, this nondivergenceless current is not associated to a continuity equation: there is no time dependence on the right-hand side, but rather a static balance equation suggesting that the nonclassical flux, acting antidiffusively, balances the classical tendency of diffusion (proportional to $\nabla^2\rho$) with the effect of stabilizing the particle-ensemble density into the exponential spatial decay associated to $q_2$ in the energy region 3.  

Performing the calculations for $x>0$ with the wave function (\ref{psievan}) in the same approximation as in the Appendix A, we obtain from the definition $\vec{u}=\tfrac{\hbar}{2m}\vec{\nabla}\ln\rho$, that $\vec{u}=u_x{\hat{i}}+u_y{\hat{j}}$, with $u_x=-\tfrac{\hbar q_2}{m}+\tfrac{\hbar q_1}{m} \tfrac{\sinh(q_1x)\chi_0(y-a)+\cosh(q_1x)\chi_0(y+a)}{\cosh(q_1x)\chi_0(y-a)+\sinh(q_1x)\chi_0(y+a)}$ \cite{notesimp}. Consider now for simplicity the limit $|\Delta| \gg \hbar J_0$, as done in \cite{sharo}. In this case $u_x\simeq -\tfrac{\hbar q_2}{m}$ (the term $q_1$ scales like $\tfrac{\hbar J_0}{|\Delta|}$), and we find, in this limit, the same relation as \cite{sharo}: $u_x=\sqrt{\tfrac{2|\Delta|}{m}}$, plotted in Fig. \ref{fig3} as a black dotted line. We remark upon, however, the following counterintuitive features of this solution: in Nelson's interpretation the velocity $u_x$ goes in the negative $x$ direction, and does not lead to density displacements (taking place only during the transient regime, as seen in Sec. II). Thus, despite obtaining the same energy-speed relation $u_x=\sqrt{\tfrac{2|\Delta|}{m}}$ as in \cite{sharo}, Nelson's picture is not quite the same as the one proposed in \cite{sharo}.

\subsection{Orthodox interpretation}

Finally, in the framework of orthodox quantum mechanics it is worth noting that the term $Q$ represents 'quantum kinetic energy fluctuations', for which no classical-like objects exist (opposite to Bohm's potential energy or Nelson's stochastic speed). In this sense the orthodox interpretation is the 'least committed' interpretation of quantum mechanics, aiming at producing correct calculations to interpret the experiments (as we have done in Sec. II and in the Appendixes), without introducing classical concepts where there might not be any. Therefore, for example, in the orthodox interpretation the exponential spatial decay of region 3 is stable because this state is an eigenstate of the Schr\"odinger equation. Its stability follows from the postulates of quantum mechanics and we should not ask for more. Of course, Fig. \ref{fig3} can be reproduced also within the orthodox interpretation, as all the calculations leading to Eq. (\ref{eqfig3}), reported in Appendix A, are based on the orthodox formalism.

\subsection{Comparison with the recent literature}

From what shown above, all interpretations (Bohm's, Nelson's and orthodox) can be used to describe Fig. \ref{fig3}. Consequently, we can affirm that the experiment of \cite{sharo} is therefore not conclusive in selecting or challenging any of these interpretations. Here we further clarify some relevant details by comparing our approach to the recent literature, in particular  \cite{drezet,drezet2,waegell,matzkin,dickau}. Daem {\it et al.} \cite{matzkin} highlight a fundamental concept that it is worth repeating here: it is impossible to falsify the Bohmian model independently of the orthodox interpretation, as they share the same mathematical structure. For this reason we did not focus on the explanations based on an experimental leakage, necessarily present in real waveguides, as highlighted, e.g., in \cite{drezet,drezet2}.
In fact, the idea of a possible contradiction between Bohm's and the orthodox approach is already suggested in the idealized theoretical model \cite{sharo2}, where the theoretical displacement of particle density in the auxiliary waveguide is deduced on the basis of the orthodox formalism, with the apparent absence of speed, $\vec{\nabla}S=0$, and in the absence of experimental leakages.
Therefore it appears crucial to analyze the challenge to Bohmian mechanics proposed by the nondissipative theoretical model of Ref. \cite{sharo2}. This necessity was already clear in \cite{dickau}, focusing on the effects of a finite wave packet to time dynamics, and, more recently, in \cite{arxiv}. Here we just remark that a wave packet has necessarily a raising edge leading to a transient regime before the stationary state sets in \cite{norsen}, of particular relevance for a gedanken experiment where the experimental leakage is absent. The (gedanken) challenge proposed in \cite{sharo2} concerns the small-time dynamics, corresponding to the effect of the raising edge alone in our approach (as shown in Appendix B, our results are not dependent on the rectangular pulse width, even in its infinite limit). So, what counts for the filling of the auxiliary waveguide is the transient time region associated to the raising edge \cite{notegauss}.
\\    
We further remark the work of Waegell \cite{waegell}, the only one discussing the role of the osmotic (Nelson's) velocity.
At the technical level, the main difference consists in the intermediate energy region $|\Delta|\leq \hbar J_0$, where his 2D numerical solution does not reproduce our 2D analytical solution (see Appendix A).
At the conceptual level, Waegell \cite{waegell} wonders if the osmotic (symmetric) velocity has a possible interpretation in terms of real motion of something: here we only remark that the approaches introduced in [1, 3] cannot answer this question. Indeed, they deal with a time reversal odd quantity (density displacement -- by reversing the time-arrow it flows in the opposite direction), whereas the answer should address a time-reversal even one (osmotic velocity).
\\
As a last remark, it is useful to remember a point raised by \cite{dickau}, who convincingly showed that the dwell times reported in \cite{sharo} for the Bohmian and orthodox interpretations are experimentally defined in two different ways, whereas the corresponding formalism suggests that their mathematical definition should be the same. Consequently, this argument of \cite{sharo} is not cogent.
\\


\section{Discussion and conclusions}

Our final picture of the experiment \cite{sharo} and of the model \cite{sharo2} is the following: when the initial wave front reaches the potential step at $x=0$, a transient regime sets in where the density profile, measured during the subsequent stationary regime, starts forming. The corresponding transient, time-dependent, wave function is described in Sec. II. During this time there exists a nonzero current density which is nondivergenceless and allows the density formation [$\partial_t \rho(x,y,t)\neq 0$] in the main and auxiliary waveguides ($w_1$ and $w_2$). 
The continuity equation (\ref{divj}) shows that the final, normalized, density $\rho_a(x)$ of  Eq. (\ref{rhoa}) measured in \cite{sharo} is formed by the time integral of the opposite of the density-current divergence. In particular, a nonzero $j_y(x,y)$ is required to transport particle density from $w_1$ to $w_2$.

Once the transient regime evolves into steady-state conditions, a divergenceless density current sets in for energy regions 1 and 2 [i.e. $\partial_t \rho(x,y,t)= 0$ everywhere], whereas the current disappears ($\vec{\nabla}S=\vec{0}$) for the evanescent wave of region 3 and the density does not change any more, in any interpretation (orthodox, Bohm's, and Nelson's). Measurable quantities in the stationary regime cannot be associated to nondivergenceless transport, so there cannot be a net density transport from $w_1$ to $w_2$ in stationary conditions. In Nelson's theoretical framework, it is possible to define a nonclassical speed and interpret Fig. \ref{fig3} in this way, though the antidiffusive character of the associated velocity makes the overall picture not equivalent to \cite{sharo}, as seen in Sec. IV. In Bohm's theoretical framework, instead, Fig. \ref{fig3} can be described as a static relation between the wave vector $q_2$ and the quantum potential $|Q|\simeq |\Delta|$, without the need of introducing a speed.
The conceptual misinterpretation in \cite{sharo} has been to associate the measured density profile at time $t$ to a density displacement at the same time $t$. Instead, by realizing that the invoked displacement from $w_1$ to $w_2$ for $x\ge 0$ takes place before, in the transient regime, we are no longer forced to think that particle density is moving at time $t$ from $w_1$ to $w_2$ and we can coherently explain the experiment in stationary conditions within Bohm ontology in terms of the relation between a static wave vector and the quantum potential \cite{note2}. This is in keeping with the general theory, clearly summarized in \cite{norsen}: Bohm's theory makes the same predictions as ordinary quantum mechanics for any experiment (at least, above the Compton wavelength \cite{bohm,note1}).

To conclude, even though, once correctly interpreted with the analysis proposed here, the challenge to Bohmian quantum mechanics disappears, we think that the experimental setup and the measurements of Sharoglazova {\it et al.} are still interesting, because they provide clear experimental data on a quantum mechanical problem with an unusual geometry (a potential step in the $x$ direction coupled to a tunnel barrier in the $y$ direction) that can be solved with analytical calculations through a relatively simple model Hamiltonian [Eq. (\ref{ham2D})], both in stationary conditions and in the transient regime. This can be of some relevance, e.g., in tunneling-based experiments, or where evanescent waves are used to create entanglement of photons \cite{ring,tun}. 
As a non-negligible add-on, it leads to a careful discussion about the meaning of i) the quantum potential in Bohm's interpretation, ii) the nonclassical diffusion velocity in Nelson's interpretation, and iii) their respective relations to the kinetic-energy fluctuations in the orthodox interpretation. This provides a complete and nontrivial example of fundamental quantum mechanics, that we think might be pedagogically useful to anybody active in the field.

\section{Acknowledgments} A fruitful discussion with Travis Norsen is gratefully acknowledged.

\section{Data availability} There are no publicly available research data or software supporting this manuscript. Requests for further information or data should be sent to the authors.

\appendix

\section{Stationary states of the bidimensional Schr\"odinger equation}

Here we propose the full stationary solution of the Hamiltonian (\ref{ham2D}), in each energy region and compare it with \cite{sharo,sharo2}. 

For $x<0$ the longitudinal eigenstates are the plane waves and the transverse eigenstate is the ground-state of the harmonic oscillator $\big(\chi_0(y)=\tfrac{1}{\sqrt{\sigma\sqrt{\pi}}}e^{-\tfrac{y^2}{2\sigma^2}}$, with  $\sigma^2=\frac{\hbar}{m\omega_0}\big)$: 
$$\psi_<(x,y)=(e^{ik_0x}+Re^{-ik_0x})\chi_0(y-a)$$ 
The total energy is $E = \frac{\hbar\omega_0}{2}+\frac{\hbar^2k_0^2}{2m}$. To compare with \cite{sharo,sharo2}, we define $E_{k_0}=E-\frac{\hbar\omega_0}{2}=\frac{\hbar^2k_0^2}{2m}$, corresponding to their total energy. In the reminder of the paper, we shall remove the constant $\frac{\hbar\omega_0}{2}$ term from the counting of the energy, as it appears for both $x<0$ and $x>0$.

For $x>0$, the transverse spectrum consists of a nearly degenerate pair of a symmetric and an antisymmetric states \cite{merz} with respect to exchange in $y\pm a$, separated by a tunnel splitting $\hbar\omega_s \simeq 2\hbar\omega_0\frac{a}{\sigma}e^{-a^2/\sigma^2} \ll \hbar\omega_0$, exponentially small for large $a$ \cite{merz}. The symmetric state $\chi_+(y)=\frac{\chi_0(y-a)+\chi_0(y+a)}{\sqrt{2(1+\Omega_S)}}$ with energy $\varepsilon_+=V_0-\frac{\hbar\omega_s}{2}$ is the ground state, whereas $\chi_-(y)=\frac{\chi_0(y-a)-\chi_0(y+a)}{\sqrt{2(1-\Omega_S)}}$ has energy $\varepsilon_- = \varepsilon_+ + \hbar\omega_s$. Here $\Omega_S=e^{-a^2/\sigma^2}$ is the overlap between $\chi_0(y-a)$ and $\chi_0(y+a)$. We have kept the overlap explicitly, though it does not play any direct role in the density measurements of \cite{sharo}, because it is necessary in order to have a nondivergenceless $j_y$ current in the transient state (see Sec. III.B). 

The global solution for $x>0$ is:

\begin{align}
\psi_>(x,y)=T_+\chi_+(y) e^{ik_+ x}+T_-\chi_-(y) e^{ik_- x}
\label{psiplus}
\end{align}

\noindent where $k_{\pm}=\sqrt{\frac{2m}{\hbar^2}(E_{k_0}-V_0\pm\frac{\hbar\omega_s}{2})}$. By imposing the continuity at $x=0$ of the function and of the derivatives, we obtain the reflection and transmission coefficients: $R=\frac{k_0-\overline{k}}{k_0+\overline{k}}$ and $T_{\pm}=\frac{2k_0}{k_0+\overline{k}}c_{\pm}=Tc_{\pm}$, with $\overline{k}=c_+^2k_++c_-^2k_-$ and $c_{\pm}=\sqrt{\frac{1\pm \Omega_S}{2}}$.

If we put $k_+=k_2+k_1$ and $k_-=k_2-k_1$ (so that $k_2=\tfrac{1}{2}(k_++k_-)$ and $k_1=\tfrac{1}{2}(k_+-k_-)$), we can write $\psi_>(x,y)$ as:

\begin{align}
\!\!\!\!\!\!\!\!\!\!\!\!\!\!\!\!\psi_>(x,y)\!=Te^{ik_2 x}\left[ \cos(k_1 x) \chi_0(y-a)\! + i \sin(k_1 x)\chi_0(y+a) \right]
\label{psisharo}
\end{align} 

This solution coincides with the solution of \cite{sharo,sharo2}, if we identify the two (abstract) degrees of freedom $\uparrow$ and $\downarrow$ associated to $\psi(x)$ in \cite{sharo,sharo2} with the transversal wave functions $\chi_0(y-a)$ and $\chi_0(y+a)$, respectively (within a phase factor). We remind that $\chi_0(y-a)$ and $\chi_0(y+a)$ are only approximately orthogonal (because of the $\Omega_S$ overlap), but to a good approximation Eq. (\ref{psisharo}) reproduces the results of \cite{sharo,sharo2}. An equivalent way to look at this equivalence is to construct directly an effective Hamiltonian from Eq. (\ref{ham2D}), by introducing, in the $x>0$ region, the transverse eigenbasis $|\chi_+(y)\rangle$ and $|\chi_-(y)\rangle$, where the one-dimensional longitudinal motion in the $x$ direction is associated to an abstract two-dimensional space spanned by the eigenbasis $|\chi_+(y)\rangle$, $|\chi_-(y)\rangle$ representing a linear combination of the two waveguides: 

\begin{align}
{\hat H}_>=-\frac{\hbar^2}{2m}\frac{\partial^2}{\partial x^2} + V_0 + \varepsilon_+ + \hbar\omega_s|\chi_-\rangle\langle\chi_-|
\label{hmaj}
\end{align}

By interpreting $\chi_0(y-a)\rightarrow \psi_{\uparrow}$ and $\chi_0(y+a)\rightarrow \psi_{\downarrow}$, we recover the full solution of \cite{sharo,sharo2} (within a phase factor and the exact orthogonality that is reached only the limit $\tfrac{a}{\sigma}\gg 1$, when, however, $\omega_s\rightarrow 0$).
We remark that, though the effective one-dimensional model \cite{sharo2} is simpler for the calculations, it may hide that the intercoupling of the two waveguides corresponds to a transverse ($y$) motion and not to an abstract degree of freedom. Such a transverse motion shows up in the transient state, where $j_y\neq 0$, as demonstrated below.

\begin{figure}[t]
\centering
\includegraphics[width=0.50\textwidth]{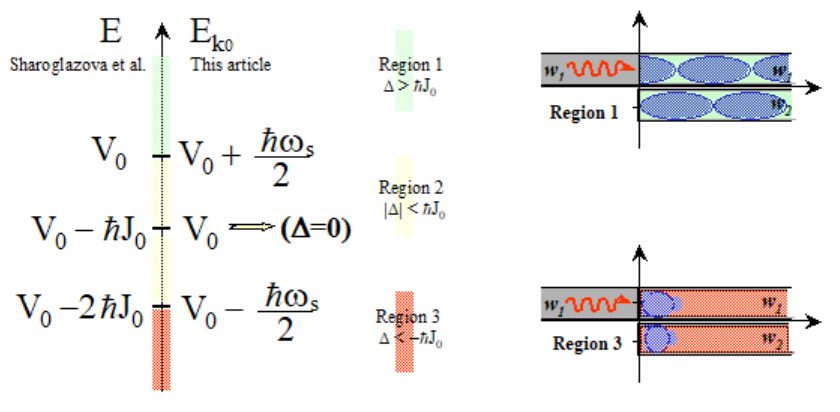} 
\caption{Left] Comparison of the energy levels in our notation and in that of \cite{sharo}. [Right] Two of the three allowed regimes described in the text. For $E_{k_0}>V_0 +\frac{\hbar\omega_s}{2}$ (region 1, upper part), both $k_+$ and $k_-$ are real and $j_x\neq 0$. The beating of the two wave vectors is responsible of the shape of the measured density in $w_1$ and $w_2$. For $E_{k_0}<V_0 -\frac{\hbar\omega_s}{2}$ (region 3, bottom diagram), both channels become evanescent. This represents the most interesting part, as the reported measurement of a speed in this regime is at the basis of the 'challenging' claim of \cite{sharo}.}
\label{fig2} 
\end{figure}

We have three different behaviours according to whether: a) $E_{k_0}>V_0 +\frac{\hbar\omega_s}{2}$ (Region 1); b) $V_0 -\frac{\hbar\omega_s}{2}<E_{k_0}<V_0 +\frac{\hbar\omega_s}{2}$ (Region 2); c) $E_{k_0}<V_0 -\frac{\hbar\omega_s}{2}$ (Region 3).
In order to compare with \cite{sharo,sharo2}, we remind that our $\frac{\hbar\omega_s}{2}$ corresponds to their $\hbar J_0$, and that their $V_0$ corresponds to the level where pure propagation starts, i.e., our $V_0 +\frac{\hbar\omega_s}{2}$. Finally, we define $\Delta=E_{k_0}-V_0$ (corresponding to the $\Delta$ defined in \cite{sharo2}). The full analogy is proposed in Fig. \ref{fig2}.

{\it Region 1.} In this case, both wave vectors are real: $k_{\pm}=\sqrt{\tfrac{2m}{\hbar^2}(\Delta\pm \hbar J_0)}$ and therefore $k_2= \frac{\sqrt{m}}{\hbar}\sqrt{\Delta+\sqrt{\Delta^2-\hbar^2J_0^2}}$ and $k_1 =\frac{\sqrt{m}}{\hbar}\sqrt{\Delta-\sqrt{\Delta^2-\hbar^2J_0^2}}$, with $\tfrac{\hbar^2k_1^2}{2m}+\tfrac{\hbar^2k_2^2}{2m}=\Delta$ and $k_1=\frac{mJ_0}{\hbar k_2}$, as in \cite{sharo2}. The latter two relations are valid in all three regions.

A current density can be associated to Eq. (\ref{psisharo}) as: $\vec{j}=\tfrac{\hbar\rho}{m}\vec{\nabla}S(x,y)$ and therefore a speed $\vec{v}=\tfrac{\vec{j}}{\rho}$. The longitudinal current is $j_x(x,y)=|T|^2\tfrac{\hbar k_2}{m}[\cos^2(k_1 x) \chi_0^2(y-a) + \sin^2(k_1 x)\chi_0^2(y+a)]+|T|^2\tfrac{\hbar k_1}{m}\chi_0(y-a)\chi_0(y+a)$, with the latter term negligible with respect to the former (proportional to the overlap of the wave functions in the two waveguides).
 So the longitudinal current is basically characterized by a transmission at the average wave vector $k_2=\tfrac{1}{2}(k_++k_-)$, with beatings at the inverse wavelength $\tfrac{2\pi}{k_1}$. These beatings are out of phase by $\frac{\pi}{2}$ in the two waveguides, so that the $x$-dependence in the auxiliary waveguide $w_2$ is $\propto\sin^2(k_1x)$. In the limit $\tfrac{a}{\sigma}\rightarrow\infty$, there would be no overlap of the two waveguides, so $\omega_s\rightarrow 0$ ($J_0\rightarrow 0$ in the notation of \cite{sharo}) and the current in the second waveguide would be zero ($k_1\rightarrow 0$).

The transverse current density is proportional to the overlap of the wave functions in the two waveguides (no overlap implies independence of $w_1$ and $w_2$, and $\chi_0(y-a)$ and $\chi_0(y+a)$ would be separately eigenstates of the global Hamiltonian): $j_y(x,y)=-|T|^2\tfrac{\hbar a}{m\sigma^2}\sin(2k_1x)\chi_0(y-a)\chi_0(y+a)$ (the negative sign shows a transfer from $w_1$ to $w_2$ for small values of $x$).
 We remark that, though $\partial_x j_x\neq 0$ and $\partial_y j_y\neq 0$ separately, the current $\vec{j}$ is divergenceless (as $\partial_x j_x=-\partial_y j_y$, within the limits of our approximation of the transverse motion, i.e. $o(\Omega_S)$) and therefore it cannot displace particle density, as obvious in stationary conditions.

{\it Region 2.} As above, $k_+$ is real but $k_-=iq_-$, with $q_-=\sqrt{\tfrac{2m}{\hbar^2}(-\Delta+\hbar J_0)}$ real. This implies that $k_2= \frac{\sqrt{m}}{\hbar}\sqrt{\Delta+i\sqrt{-\Delta^2+\hbar^2J_0^2}}=k_1^*$. 

The wave function is given by: $\psi_>(x,y)=T \allowbreak \left[\chi_0(y-a)(e^{ik_+ x}+e^{-q_- x})+\chi_0(y+a)(e^{ik_+ x} - e^{-q_- x})\right]$. The current in the longitudinal direction is: $j_x(x,y)=\tfrac{\hbar}{m}|T|^2  \big[k_+c_+^2\chi_+^2 +e^{-q_-x}(k_+\cos(k_+x)+q_-\sin(k_+x)) \allowbreak (\chi_0^2(y-a)-\chi_0^2(y+a))/4\big]$. The transversal current is: $j_y=\tfrac{\hbar a}{m \sigma^2}|T|^2e^{-q_-x}\sin(k_+x)\chi_0(y-a)\chi_0(y+a)$. We remark that $j_y$ only depends on the overlap of $\chi_0(y-a)$ and $\chi_0(y+a)$.
Again, as in the previous case, though $\partial_x j_x\neq 0$ and $\partial_y j_y\neq 0$, the current density $\vec{j}$ is divergenceless, as it should be.

{\it Region 3.} This is the critical region, as it 'challenges' Bohmian mechanics. In this case, both wave vectors $k_+=iq_+$ and $k_-=iq_-$ are imaginary and there is no stationary current. We have $q_{\pm}=\sqrt{\tfrac{2m}{\hbar^2}(-\Delta\mp\hbar J_0)}$, with $q_->q_+>0$. We obtain therefore: $q_{1,2}=\frac{\sqrt{m}}{\hbar}\sqrt{-\Delta\mp\sqrt{\Delta^2-\hbar^2J_0^2}}$ ($-$ for $q_1$ and $+$ for $q_2$). We remark that, when $|\Delta|\gg \hbar J_0$, $q_1\rightarrow 0$ and $q_2\rightarrow \tfrac{2m}{\hbar^2}|\Delta|$. The wave function in Region 3 is:

\begin{align}  \label{psievan}
\!\!\!\!\!\!\!\!\!\!\!\!\!\!\!\!\psi_> & (x,y)=T\left(c_+\chi_+(y) e^{-q_+ x}+c_-\chi_-(y) e^{-q_- x}\right) \\
&\!\!\!\!\!\!\!\!=Te^{-q_2 x}\left[ \cosh(q_1 x) \chi_0(y-a) + \sinh(q_1 x)\chi_0(y+a) \right] \nonumber
\end{align}

\section{Transient regime in the one- and two-dimensional cases}

It might be useful here to remind the main outcomes of the one-dimensional case of a step potential, in transient conditions, along the description of, e.g., \cite{review,moshinsky}.
Consider the one-dimensional Schr\"odinger equation in a step potential:

\begin{equation}
i\hbar\frac{\partial\psi}{\partial t}=\left[-\frac{\hbar^2}{2m}\frac{\partial^2}{\partial x^2} +V_0\theta(x)\right]\psi
\end{equation}
where $V_0$ is a positive constant and $\theta(x)$ the Heaviside function (1 if $x>0$ and 0 otherwise). 
Define as usual $k=\frac{\sqrt{2mE}}{\hbar}$ and $\kappa=\frac{\sqrt{2m(V_0-E)}}{\hbar}$.
Then, take a rectangular wave packet, with $E<V_0$, at the initial time $t=0$: $\psi(x,0)=e^{ik_0x}\left(\theta(x+L)-\theta(x)  \right)$, with $L\gg 1/k_0$. This wave packet is not strictly monochromatic at $k_0$ because of the cutting edge at $x=0$ (the one at $x=-L$ is not actually necessary in our case, as we are interested only in the transient phenomena at the raising edge: it will be eliminated below by letting $L\rightarrow \infty$). The packet travels towards the positive $x$ direction and interacts with the step potential, at $x=0$, for $t\ge 0$. 

The solution of $\psi(x,t)$ for $x>0$, $t>0$ can be obtained using the retarded Green function $G_t(x,x';t)$ by standard techniques \cite{review,moshinsky,moshinsky2}, as detailed below. Start from the stationary, energy-dependent, Green function $G_E(x,x';E)=\frac{m}{i\hbar^2k}T(k)e^{-ikx'}e^{-\kappa(k)x}$, valid for $x'<0$ and $x>0$. Here $T(k)=\frac{2k}{k+i\kappa(k)}$is the transmission coefficient and $\kappa(k)=\sqrt{\frac{2mV_0}{\hbar^2}-k^2}$. Then Fourier-transform to $G_t(x,x';t)=\int_{-\infty}^{+\infty} G_E(x,x';E)e^{-iEt/\hbar}\frac{dE}{2\pi}$. Transform the $E$-integral in a $k$-integral by $E=\frac{\hbar^2k^2}{2m}$, $dE=\hbar^2kdk/m$ and approximate $T(k)\simeq T(k_0)$ (slowly varying around the pole $k=k_0$). In order to perform the integral, we should also approximate \cite{technote} $\kappa(k)\simeq \kappa_0+\frac{k_0}{\kappa_0}(k-k_0)$, where $\kappa_0=\kappa(k_0)$. Then the integral can be formally calculated and we find:

\begin{align} \label{erfc}
\!\!\!\!\!\! \psi(x,t) &=\int_{-\infty}^{+\infty}dx' \psi(x',0)G_t(x,x';t)\\
& =\int_{-L}^0 dx' e^{ik_0x'}G_t(x,x';t) \nonumber \\
& \simeq \frac{1}{2}T(k_0) e^{-\kappa_0 x}e^{-iE_0t/\hbar}  {\rm erfc}\left(z(x,t) \right) \nonumber
\end{align}

\noindent with $z(x,t)\!=\!-\frac{k_0}{\kappa_0}\sqrt{\frac{m}{2\hbar t}}\left(x - iv_0t\right) e^{i\pi/4}$ and $v_0=\frac{\hbar \kappa_0}{m}$.
We remind that $\frac{1}{2} {\rm erfc}\left(z(x,t) \right) \xrightarrow[t\rightarrow +\infty]{ } 1$, so that we recover in this limit the usual stationary solution $\psi_{\rm st}(x,t)=T(k_0) e^{-\kappa_0 x}e^{-iE_0t/\hbar}$, characterized by a null spatially-dependent phase ($\partial_x S=0$) and no current. Yet, this is not true during the transient period, when $\psi(x,t)$ has a spatially-dependent phase, contained in the complementary error function erfc. As it is numerically \cite{notenum} shown in Fig. 20 of \cite{review}, the time-dependence of $|\psi(x,t)|^2$ leads to a diffusion current and not a propagating current: the density probability does not grow 'from left to right' in $x$, as we would have had if the argument of the error function had been $x-v_0t$ (i.e. for $E>V_0$)). Instead, the imaginary unit in the expression $x-iv_0t$ is the signal of a spatially-weighted diffusion ('from bottom to top') that is the consequence of the parabolic character of the Schr\"odinger equation, leading to the instantaneous diffusion (with infinite velocity) of all the components of the initial wave packet. This deficiency might be cured describing the system with the relativistic Klein-Gordan equation, allowing a finite propagation speed through a $\theta(ct-x)$ in the Green function (as, for $c\rightarrow \infty$, $\theta(ct-x)\rightarrow 1$, one recovers the nonrelativistic instantaneous increase of the density throughout the whole $x$-axis \cite{review}). Regardless, had we even chosen a different profile (e.g. gaussian) for the initial wave function, the main message would not have changed: during the transient period before the stationary state, there is a density current whose nonzero divergence allows building up the probability density of the evanescent wave. Once the stationary state sets in, the wave function becomes real and the probability current becomes identically zero for an evanescent state. The evanescent density profile at $x>0$ during the stationary state is therefore the result of the above nonzero transient current converging towards an eigenstate of the stationary Schr\"odinger equation.

The generalization to two dimensions is straightforward. We first evaluate the energy-dependent Green function, that obeys \cite{fetter} $(E-\hat{H})G_E(x,x';y,y')=\delta(x-x')\delta(y-y')$. As we need to evaluate the transition of a particle initially in the negative-$x$ region and in the main waveguide ($w_1$) towards the positive-$x$ region and the auxiliary waveguide ($w_2$), we take $x'<0$, $y'>0$, $x>0$ and any $y$.
The Green function takes the form \cite{fetter} $G_E(x,x';y,y')=\Sigma_{\alpha=\pm}\chi_{\alpha}(y)\chi_{\alpha}(y')g_{\alpha}(x,x';E)$, where $\chi_{\alpha}(y)$ is one of the two transverse solutions (reported in Appendix A) and $g_{\pm}(x,x';E)$ is the solution of the one-dimensional, channel-dependent, Green function equation:  $\left[E_{k_0} +\frac{\hbar^2}{2m}\partial_x^2 -E_{\pm} \right]g_{\pm}(x,x';E)=\delta(x-x')$, with $E_{\pm}=V_0\mp \frac{\hbar\omega_s}{2}$, as $x>0$ (see Fig. \ref{fig2} in the Appendix A for the definitions of the energy terms). 

In the evanescent region $E_{k_0}<V_0-\frac{\hbar\omega_s}{2}$, and the solution is: $g_{\pm}(x,x';E)=\frac{m}{i\hbar^2 k} T_{\pm}(k)e^{ikx'}e^{-q_{\pm}(k) x}$, like in the one-dimensional case (reported in Appendix B), except for the channel dependence.
Here $T_{\pm}(k)=\frac{2k}{k+i\overline{q}}$, with $\overline{q}=q_+c_+^2+q_-c_-^2$ and $q_{\pm}(k)=\sqrt{\frac{2m}{\hbar^2}\left(V_0-E_{k_0}\mp\frac{\hbar\omega_s}{2} \right)}$.

Therefore, the general solution for the wave function is obtained by summing over the two one-dimensional Green functions (one for each transverse eigenchannel), and with the same hypothesis used in the Appendix B for the one-dimensional case, we get:

\begin{align}
\!\!\!\!\!\!\!\!\!\!\!\!\!\!\!\psi(x,y,t)=\frac{e^{-\frac{iE_0t}{\hbar}}}{2} \!\!\! \sum\limits_{\alpha=\pm}\!c_{\alpha}\chi_{\alpha}(y) T_{\alpha}(k_0)e^{-q_{\alpha}^{(0)}x} {\rm erfc}[z_{\alpha}(x,t)],
\label{psi2D}
\end{align}

\noindent where $z_{\alpha}(x,t)=-\frac{k_0}{q_{\alpha}^{(0)}}\sqrt{\frac{m}{2\hbar t}}\left(x - iv_{0\alpha}t\right) e^{i\pi/4}$, and $v_{0\alpha}=\frac{\hbar q_{\alpha}^{(0)}}{m}$ have the same expression as for a single channel. In particular, as in that case, $\frac{1}{2} {\rm erfc}\left(z_{\alpha}(x,t) \right) \xrightarrow[t\rightarrow +\infty]{ } 1$, so that we recover in this limit the stationary solution $\psi(x,y,t)=\frac{e^{-\frac{iE_0t}{\hbar}}}{2} \!\!\! \sum\limits_{\alpha=\pm}\!c_{\alpha}\chi_{\alpha}(y) T_{\alpha}(k_0)e^{-q_{\alpha}^{(0)}x}$ of Eq. (\ref{psiplus}). We remark that the total energy $E=E_{k_0}+\frac{\hbar \omega_0}{2}$ in Eq. \ref{psi2D} is just an overall phase factor, as it should be for a truly stationary condition, and therefore does not contribute to the density nor to the current: all the time-dependence comes from the complementary error function ${\rm erfc}(z_{\alpha}(x,t))$, whose channel dependence $\alpha$ leads to cross contributions in the calculation of the total density and of the density current.


\begin{thebibliography}{99}
\bibitem{note1}
In the Hamiltonian formalism of quantum mechanics kinetic and potential energies appear in a symmetric way and it is not necessarily evident how to identify each. For example, potential and kinetic energy can be inverted by a canonical transformation in the harmonic oscillator and, in one-dimensional radial problems, the kinetic origin of the centrifugal barrier in the orthoradial direction cannot be guessed without further information. It is the same idea that led Bohm to the interpretation of the average kinetic-energy fluctuations of orthodox quantum mechanics, $\frac{\hbar^2}{8m}\frac{(\vec{\nabla}\rho)^2}{\rho^2}$, as a potential energy - the quantum potential. According to Bohm \cite{bohm}, the two interpretations could be disentangled only with sub-Compton-wavelength measurements. In the case of \cite{sharo}, it would mean below $\simeq 50$ nm.
\bibitem{notesimp} 
We are interested in the component $u_x$ as it is the quantity measured in \cite{sharo} through the procedure of Eq. (\ref{rhoa}), that referred to the speed in the $x$-direction.
\bibitem{notegauss}
A rectangular or a gaussian wave packet are conceptually equivalent: we used a rectangular wave packet in order to push as far as possible the analytical calculations with Green function and error function. In \cite{norsen} the choice of a more realistic gaussian pulse was adopted. We remark that also in Fig. 4 of \cite{norsen}, the filling of the auxiliary waveguide takes place during the raising edge of the gaussian pulse, whereas the trailing edge of the pulse acts as a sink, emptying the auxiliary waveguide.
\bibitem{note2}
We remark that \cite{drezet} has already criticized the operational definition of a speed as defined in \cite{sharo, sharo2}.
\bibitem{notenum} 
Analytically, we might just demonstrate the presence of a nondivergenceless current in some specific limits, as, for example, for $t\rightarrow 0^+$, when $z(x,t)\rightarrow -\sqrt{\frac{m}{2\hbar t}} \frac{k_0}{\kappa_0}x e^{i\pi/4}$. Keeping only the leading term of the asymptotic expansion of erfc(z): ${\rm erfc}(z) \xrightarrow[|z|\rightarrow +\infty]{ } \frac{e^{-z^2}}{\sqrt{\pi}z} +o(|z|^{-1})$, we obtain (for $x>0$) the time-dependent probability density (remark that $z^2$ is purely imaginary and equal to 1 in the $t=0^+$ limit) $|\psi(x,t=0^+)|^2\simeq \frac{1}{\pi}|T(k_0)|^2 e^{-2\kappa_0 x} \frac{\kappa_0}{k_0} \frac{\hbar  t}{2m  x^2}$. From the continuity equation we get in this limit a nonzero, nondivergenceless current.
\end{thebibliography}
\end{document}